\documentclass[10pt,letterpaper]{article}
\usepackage[top=0.85in,left=2.75in,footskip=0.75in,marginparwidth=2in]{geometry}

\usepackage[utf8]{inputenc}

\usepackage{cite}

\usepackage{nameref,hyperref}

\usepackage[right]{lineno}

\usepackage{microtype}
\DisableLigatures[f]{encoding = *, family = * }

\raggedright
\usepackage{changepage}

\usepackage{graphicx}%
\usepackage{multirow}%
\usepackage{amsmath,amssymb,amsfonts}%
\usepackage{amsthm}%
\usepackage{mathrsfs}%
\usepackage[title]{appendix}%
\usepackage{xcolor}%
\usepackage{textcomp}%
\usepackage{manyfoot}%
\usepackage{booktabs}%
\usepackage{algorithm}%
\usepackage{algorithmicx}%
\usepackage{algpseudocode}%
\usepackage{listings}%
\usepackage{comment}
\usepackage{caption}
\usepackage{soul}

\usepackage{geometry}
\newcommand{\ket}[1]{|#1\rangle}

\usepackage[aboveskip=1pt,labelfont=bf,labelsep=period,singlelinecheck=off]{caption}

\makeatletter
\renewcommand{\@biblabel}[1]{\quad#1.}
\makeatother

\usepackage{lastpage,fancyhdr,graphicx}
\usepackage{epstopdf}
\fancyheadoffset[L]{2.25in}
\fancyfootoffset[L]{2.25in}

\usepackage{color}

\definecolor{Gray}{gray}{.25}

\usepackage{graphicx}

\usepackage{sidecap}

\usepackage{wrapfig}
\usepackage[pscoord]{eso-pic}
\usepackage[fulladjust]{marginnote}
\reversemarginpar

\begin{document}
\vspace*{0.35in}

\begin{flushleft}
{\Large
\textbf\newline{End-to-End Demonstration of Quantum Generative Adversarial Networks for Steel Microstructure Image Augmentation on a Trapped-Ion Quantum Computer}
}
\newline
\\
Samwel Sekwao\textsuperscript{1,*},
Jason Iaconis\textsuperscript{1},
Claudio Girotto\textsuperscript{1},
Martin Roetteler\textsuperscript{1},
Minwoo Kang\textsuperscript{2},
Donghwi Kim\textsuperscript{2},
Seunghyo Noh\textsuperscript{2},
Woomin Kyoung\textsuperscript{2},
Kyujin Shin\textsuperscript{2,\textdagger}
\\
\bigskip
\bf{1} IonQ Inc., College Park, MD, 20740, USA
\\
\bf{2} Materials Research \& Engineering Center, Advanced Vehicle Platform Division, Hyundai Motor Company, Uiwang, 16082, Republic of Korea
\\
\bigskip
\textsuperscript{*} samwel.sekwao@ionq.co\\
\textsuperscript{\textdagger}  shinkj@hyundai.com

\end{flushleft}

\section*{Abstract}
Generative adversarial networks (GANs) are a machine learning technique capable of producing high-quality synthetic images. In the field of materials science, when a crystallographic dataset includes inadequate or difficult-to-obtain images, synthetic images can be used for image augmentation to mitigate data scarcity and streamline the preparation of datasets for high-throughput analysis. We integrate quantum computing with GANs into a hybrid quantum-classical GAN to generate complex 5-channel electron backscatter diffraction (EBSD) images of two distinct microstructure phases of steel. 

By training a quantum circuit at the input layer of a large classical Wasserstein GAN (WGAN) model, we achieve higher image quality compared to a baseline classical GAN. The choice of WGAN also helps to mitigate mode collapse. We generate images from both ferrite and bainite microstructure phases in an end-to-end workflow. 
With respect to maximum mean discrepancy score, we find that the hybrid quantum-classical WGAN improves over classical Bernoulli GANs in 70\% of samples. 

As the quantum computer is part of the training procedure, our method has potential to scale to larger number of qubits. Our results indicate that the WGAN model based on the quantum circuit ansatz may be effectively leveraged to enhance the quality of synthetic EBSD images on both quantum simulators and actual quantum hardware.


\section*{Introduction}\label{sec1}

The field of quantum machine learning (QML) has seen significant interest in recent years. A vast body of literature has emerged, demonstrating the potential of training parameterized quantum circuits to learn from data \cite{Biamonte_2017,PhysRevA.98.032309,farhi2018classificationquantumneuralnetworks,PhysRevA.101.032308,Cerezo_2021,Hur_2022}. While there have been some efforts to understand the practical aspects of applying these methods to real world problems \cite{cherrat2024quantum, thakkar2024improved, hibat2024framework, iaconis2024quantum}, much of the existing work has been limited to either highly simplified problems or to primitive datasets where classical artificial intelligence (AI) methods already excel. In the field of generative modeling, quantum circuits are able to sample from probability distribution which are provably hard to generate using classical methods \cite{zhu2022generative,harrow2017quantum,farhi2016quantum,boixo2018characterizing}. When applied to the task of modeling classical probability distributions, these models have shown evidence of improved expressivity and generalization capabilities, leading  to superior models when evaluated on a parameter-by-parameter basis \cite{zhu2022generative,gao2022enhancing,gili2024generalization,caro2022generalization}. On the other hand, QML models have been limited in their ability to challenge modern classical generative machine learning (ML) models due to the limited qubit count and circuit depth available on modern devices in the noisy intermediate-scale quantum (NISQ) era \cite{Preskill_2018}. Most quantum devices available to date are limited to a $\mathcal{O}(10^2-10^3)$ qubit count and can only execute a few hundred two-qubit gate operations before noise overshadows the computation.  One potential path forward is through the hybrid integration of both classical and QML models. In particular, one may include a quantum circuit as a component within a much larger hybrid quantum-classical ML model.

The hybrid integration approach, introduced in Ref.~\cite{PhysRevX.12.031010}, leverages a  parameterized quantum circuit to generate a correlated input distribution of binary strings. This distribution serves as the random latent vector input for a larger classical generative adversarial network (GAN) model. In contrast, the standard GAN implementation introduces  randomness to the model by sampling a latent vector from a standard multi-dimensional probability distribution. The quantum approach is based on the associative adversarial network GAN (AAN-GAN) model. By sampling the input latent vector from a trained correlated distribution, the model performance can be improved \cite{RBM_AAN,Alcazar_2020}.  In particular, the input distribution is trained to mimic the distribution of the penultimate output layer of the discriminator portion of the GAN, sampled over the set of real and generated images.
In \cite{PhysRevX.12.031010}, the authors demonstrated that a quantum circuit can effectively generate such correlated distribution over multi-dimensional binary strings, producing images with improved quality over the standard GAN approach for generating images of handwritten digits. 

In this work, we improve and expand upon the hybrid quantum-classical approach to the GAN model in several key directions. First, we train the quantum layer of the hybrid model to directly optimize the GAN loss function, as opposed to matching the output of the discriminator layer. Second, we apply this hybrid quantum-classical GAN model to a much more complicated data set. To accommodate this significant leap in complexity, we adapt the classical model to incorporate the recent advancements in the base GAN approach. To this end, we apply the Wasserstein GAN (WGAN) methodology to generate synthetic set of complex 5-channel electron backscatter diffraction (EBSD) images from different phases of steel microstructures.  We would like to stress that these are not simply superficial differences. We will demonstrate that there are inherent limitations to integrating the quantum AAN-GAN routine to more complex hybrid models, and we suggest that the improvement observed in the previous works may have partly resulted from the simplicity of the classical model. Our implementation solves these issues and integrates the quantum and classical models tightly, making them inseparable into different components. In this paper, we will describe our improved hybrid approach to the quantum-classical GAN model and through numerical experiments we demonstrate that the inclusion of a quantum layer consistently improves over the best classical results. We achieve this through classical simulations of the quantum circuit layer and actual execution of our algorithm on a trapped-ion quantum computer. We validate the robustness of our method by training the entire model end-to-end on the actual quantum hardware. Overall, we believe that our new implementation provides a path towards scaling this hybrid quantum-classical GAN methodology to state-of-the-art levels. Quantum Wasserstein GANs have been used in other application contexts, including option pricing~\cite{Fuchs:2023}, high-energy physics~\cite{Baglio:2024}, materials science~\cite{Jurasz:2023}, anomaly detection~\cite{Herr:2010}, and theoretical computer science~\cite{Chakrabarti:2019}. 

The rest of this paper is structured as follows. In section 2, we present a brief introduction to GANs, WGANs, and quantum-circuit Born machines (QCBMs). We also describe the methodology used to train quantum enhanced WGANs, and introduce the metric used to evaluate GANs/WGANs trained on image data. Section 3 introduces the back-scatter diffraction images used to train the classical and quantum models. In section 4, we compare the performance of the classical models to the quantum enhanced models trained on both an ideal simulator and IonQ's Aria-2 quantum processing unit (QPU) and describe briefly the experimental setup.

\section*{Methods}

\subsection*{Generative Adversarial Networks (GANs)}\label{sec2}
GANs \cite{NIPS2014_5ca3e9b1} are one of the most popular forms of generative ML models in use today. GANs have shown a significant ability to produce highly realistic images and have also been applied to generate other types of synthetic data \cite{radford2016unsupervisedrepresentationlearningdeep,brock2019largescalegantraining,Yi2018GenerativeAN,10.1093/mnrasl/slx008,de_Oliveira_2017}.
The prototypical GAN comprises two components: a generator ($G$) and a discriminator ($D$), both implemented as artificial neural networks. The generator’s objective is to generate synthetic images from a random input sampled from a latent space distribution. Conversely, the discriminator’s goal is to accurately distinguish between real images from a training dataset and synthetic images produced by the generator. The generator aims to deceive the discriminator by producing high-quality samples that the discriminator cannot distinguish from real samples. In the case of deep convolutional GAN (DC-GAN), which is commonly used when training on image datasets, both the generator and discriminator components are deep convolutional neural networks. These two networks compete to find the equilibrium of the GAN adversarial-loss given by
\begin{equation}\label{eqn:1}
\min_{G}\max_{D} \underset{\mathbf{x}\sim\mathbb{P}_r}{\mathbb{E}}[\log(D(\mathbf{x})] + \underset{\Tilde{\mathbf{x}}\sim\mathbb{P}_g}{\mathbb{E}}[\log(1 - D(\Tilde{\mathbf{x}})],
\end{equation}
where $\mathbf{x}$ is a sampled from the distribution of real data, $\mathbb{P}_r$, and $\Tilde{\mathbf{x}} = G(z)$  sampled from the synthetic data distribution, $\mathbb{P}_g$, produced by the generator. Here, $z$ is a latent vector that is input to the generator and is sampled from some prior distribution, $\mathbb{P}_z$. Typically, $\mathbb{P}_z$ is a multidimensional uniform distribution of dimension $n_z$. However, it has also been shown that $z$ can be sampled from a $n_z$ dimensional binary distribution \cite{brock2019largescalegantraining}.

Despite their success in generating high quality data, there are still many challenges associated with training GANs. Most notoriously, the training often suffers from mode collapse and instability \cite{NIPS2014_5ca3e9b1,brock2019largescalegantraining}. The difficulty arises from the delicate zero-sum game between the generator and discriminator to find the equilibrium for Eq.~\ref{eqn:1} \cite{arjovsky2017principledmethodstraininggenerative}.
WGANs are variants of GANs that were introduced to improve GAN training stability and avoid mode collapse \cite{arjovsky2017wassersteingan}. In WGANs, the loss function of Eq.~\ref{eqn:1} is modified to instead minimize the \textit{Earth-Mover} distance (\textit{i.e.}, Wasserstein-1 distance) between the distributions of real and generated images \cite{arjovsky2017wassersteingan}. The new WGAN adversarial-loss equation is given by 

\begin{equation}\label{eqn:2}
\min_{G}\max_{D} \underset{\mathbf{x}\sim\mathbb{P}_r}{\mathbb{E}}[D(\mathbf{x})] - \underset{\Tilde{\mathbf{x}}\sim\mathbb{P}_g}{\mathbb{E}}[D(\Tilde{\mathbf{x}})],
\end{equation}
where the discriminator $D$ belongs to a set of 1-Lipschitz functions. In order to enforce the Lipschitz constraint, the weights of $D$ are clipped within the range $[-c,c]$\cite{arjovsky2017wassersteingan}. This weight clipping can lead to either vanishing or exploding gradients if the parameter $c$ is not chosen carefully and more stable gradients can be obtained by implementing a gradient penalty on the discriminator norm to enforce the Lipschitz constraint \cite{gulrajani2017improvedtrainingwassersteingans}. The adversarial-loss in this case becomes

\begin{equation}\label{eqn:3}
\min_{G}\max_{D} \underset{\Tilde{\mathbf{x}}\sim\mathbb{P}_g}{\mathbb{E}}[D(\Tilde{\mathbf{x}})] -\underset{\mathbf{x}\sim\mathbb{P}_r}{\mathbb{E}}[D(\mathbf{x})]  + \lambda \underset{\hat{x}\sim P_{\hat{x}}}{\mathbb{E}}[(||\nabla_{\hat{x}}(D(\hat{x})||_2-1)^2],
\end{equation}
where $\lambda$ is the gradient penalty parameter, and $\hat{x} = \epsilon x + (1-\epsilon)\tilde{x}$ is the mixture of real and generated images ($\epsilon \in [0,1]$). 

\subsection*{Quantum-Circuit Born Machine (QCBM)}

The field of QML explores the potential of quantum computers to improve classical AI algorithms. 
 By training a quantum wave function using classical data, the quantum computer encodes a target probability distribution, whose elements can be sampled with probability given by the Born rule by measuring the quantum state. Quantum circuits are capable of efficiently generating probability distributions that cannot be efficiently reproduced classically. By increasing the set of potential distributions accessible to a generative ML model, one might therefore expect that these quantum distributions can be used to improve the performance of some generative modeling tasks.

QCBMs are a type of quantum generative models where a variational quantum circuit is used to generate a class of quantum states controlled by a set of variational parameters $\theta$. For a given instance of the parameters $\theta$, the state of the QCBM is given by the wave function:

\begin{equation}\label{eqn:4}
|\Psi(\theta)\rangle = U(\theta)|\vec{0}\rangle = \sum_x c_x(\theta) \ket{x},
\end{equation}
where $U(\theta)$ is a parameterized quantum circuit, $|\vec{0}\rangle$ is the initial state of the QCBM, and $\ket{x}$ is one of the $2^N$ computational basis states with associated binary bitstring $x$. The variational parameters, $\theta$, may be tuned to minimize a loss function based on a training data set. 

\begin{figure}[H]
    \centering
    \includegraphics[width=0.7\linewidth]{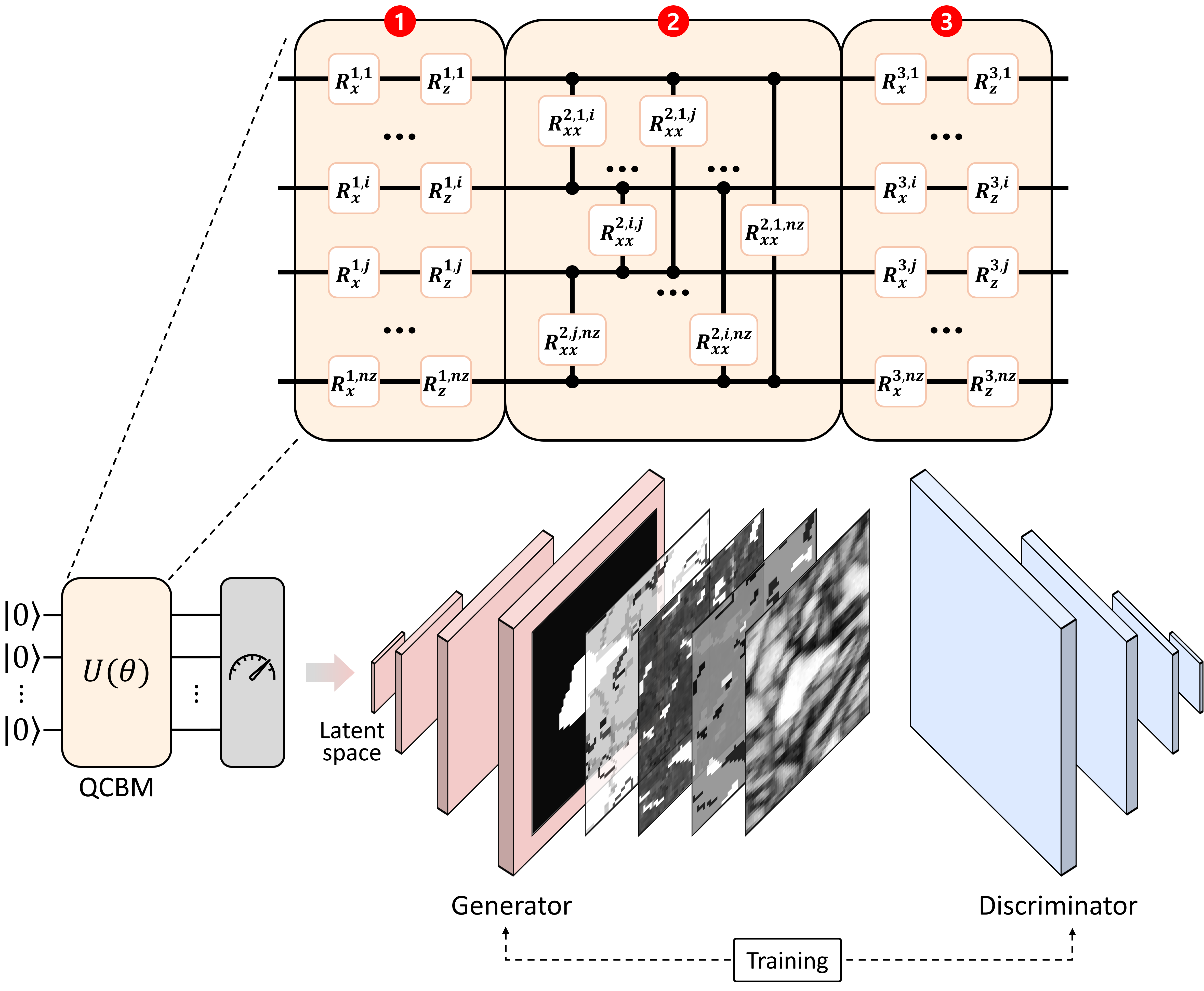}
    \caption{A schematic diagram of the full QCBM-WGAN workflow explored in this work. The sampled output of the QCBM is used as input to the classical WGAN model. Both components are trained together to produce 5-channel EBSD images. In the quantum circuit ansatz used for the QCBM, layers 1 and 3 consist of single-qubit rotations, while second layer is composed of two-qubit $R_{xx}$ gates between all pairs of qubits. Each gate has a separate variational parameter that can be independently optimized.}
    \label{fig:QCBM_WGAN}
\end{figure}

The quantum circuit for $U(\theta)$ used to implement the QCBM in this work (Fig. \ref{fig:QCBM_WGAN}) consists of three regions: regions 1 and 3 with alternating $R_\textit{x}$, $R_\textit{z}$ gates, and region 2 with $R_\textit{xx}$ gates with all-to-all connectivity. This one layer configuration can be expanded with additional layers to increase expressivity by adding the gates in regions 2 and 3 for each additional layer.

For a QCBM with $n$ qubits, there are $2^n$ computational basis states $|0\rangle...|2^{n-1}\rangle$. The probability distribution $q_{\theta}$ is modeled using the Born probabilities given by
\begin{equation}\label{eqn:4}
q_{\theta}(j) = |\langle j|U(\theta)|0\rangle|^2
\end{equation}
where $j\in[0,2^n-1]$ \cite{PhysRevX.12.031010}.

\subsection*{QCBM Enhanced WGANs}
As mentioned earlier, the $n_z$-dimensional input vector $z$ that is fed into the GAN generator is typically sampled from a relatively simple prior distribution, such as a uniform or normal distribution. Recent work showed that $z$ can alternatively be sampled from a discrete Bernoulli distribution,
$z \in \{0,1\}^{\bigotimes N}$, and generate high quality synthetic data
\cite{brock2019largescalegantraining}. We leverage this solution for our fully classical WGANs, which we refer to as
\textit{Bernoulli} WGANs, sampling $z$ from a $n_z$ dimensional Bernoulli distribution with $p=0.5$.

There is evidence that the performance of different WGAN structures and datasets can be improved if the latent vector is sampled from a distribution that is trained along with the generator and discriminator \cite{PhysRevX.12.031010}. One method for training a nontrivial input latent vector distribution is with ANNs. The approach can be extended to include a quantum component, where the latent vector is sampled from the output of a trained QCBM\cite{PhysRevX.12.031010,Alcazar_2020}. This QCBM-AAN hybrid quantum classical model has a structure like that shown in Fig.~\ref{fig:QCBM_WGAN}.

In QCBM-AANs, the latent vector distribution $q_{\theta}(z)$ encoded in the QCBM is trained to maximize its likelihood to the distribution of the outputs of the latent layer of the discriminator $p_l(z)$\cite{PhysRevX.12.031010}. Specifically, the output distribution of the QCBM is trained to minimize the Kullback–Leibler (KL) divergence between $q_{\theta}(z)$ and $p_l(z)$. However, as $n_z$ becomes large, calculating the KL divergence becomes intractable as the support of these distributions grows exponentially large in $n_z$. In addition to this, in order to minimize this KL divergence, the penultimate output layer of the discriminator, $p_l(z)$, must be converted to a $n_z$-dimensional binary distribution to match the format of $q_\theta(z)$. This conversion generally requires setting a custom threshold, where $[p_l(z)]_i$ is set to zero (one) when $z_i$ is below (above) this threshold. However, this thresholding procedure is rather ad-hoc, can become challenging as the optimal threshold parameter is not known beforehand, and can lead to abnormal behavior in certain cases.  

To circumvent these difficulties, we implemented a new training method whereby the QCBM parameters, $\theta$, are optimized to minimize the full WGAN generator loss, according to:
\begin{equation}
L_G(\theta) = -D[G(\theta)],
\label{eqn:Gen_Loss}
\end{equation}
where $G(\theta)$ is a batch of images generated by the discriminator and $D[G(\theta)]$ is the average discriminator output for the batch. The parameters $\theta$ can be optimized by using gradient descent, by updating 
\begin{equation}
\theta:= \theta - \alpha \nabla_{\theta}L_G(\theta),
\end{equation}
where $\alpha$ is the learning rate. 

Typically, the gradient with respect to the quantum circuit parameters can be calculated using by back-propagation through the classical GAN model. The derivatives with respect to $\theta$ can often be calculated using the so-called parameter shift rules \cite{mitarai2018quantum,schuld2019evaluating} when an observable measured at the output of the quantum circuit. However, this procedure is not possible in our case, as we instead sample random bitstrings from the  output distribution rather than measure a simple observable.  An alternative strategy is to calculate the gradient using a discrete finite-difference approximation.

The gradients $\nabla_{\theta}L_G(\theta)$ are computed numerically as in
\begin{equation}
\nabla_{\theta}L_G(\theta) \approx \frac{L_G(\theta + \delta) - L_G(\theta-\delta)}{2\delta}.
\label{eqn:Gloss_Grad}
\end{equation}

The computation requires the evaluation of two quantum circuits for each parameter that is updated. In some circumstances, it is important to limit the total number of quantum circuit evaluations needed to train the GAN model. One way to accomplish this is by using a gradient-free optimization method.   The simultaneous perturbation stochastic approximation (SPSA) algorithm \cite{SPSA}, is such an alternative approach, which we use for training the QCBM on the real quantum hardware. 
In the SPSA training framework, the relevant QCBM hyper-parameters are the number of qubits, $n_z$, the numerical gradient parameters, $\alpha$ and $\delta$, and the number of images used to estimate the gradients, $N_\textit{samples}$.

The 1-qubit, and 2-qubit gate speeds for Aria-2 QPU are 135$\mu$s and 600$\mu$s respectively\cite{AriaGateSpeeds}. As a result, sampling from the QPU is relatively slow. A strategy to speed up the overall process is to separate the training of the QCBM from the training of the WGAN. In details, we fix the value of $\theta$ and optimize the classical parameters for $m$ optimization steps; we then freeze the classical parameters and update the parameters $\theta$ of the QCBM. The update frequency $m$ is an additional hyperparameter for our experiments. To ensure stability in the WGAN training we only train the QCBM model for the first $M$ epochs, and freeze the parameters for the remainder of the training \cite{PhysRevX.12.031010}.  

\subsection*{Evaluation Metrics}

Evaluating the performance of generative models is a complex task with many considerations affecting the quality and suitability of a given metric \cite{borji2022pros, xu2018empirical}. Standard methods for evaluating image quality in generative models often leverage high quality image classification models that have been separately trained using supervised ML. The inception score (IS) or the Frechet inception distance (FID) are common examples of such metrics in literature, which both rely on the availability of a separate high quality classification model. It is difficult to apply these metrics to bespoke image data sets with limited data, where the training of such a high quality classification model may be challenging. One choice of an alternative metric which does not require a pre-trained classification model is the maximum mean discrepancy (MMD) score. This metric  measures the similarity between two probability distributions using an expectation value of the distance between the real and generated imaged, and is given by:

\begin{eqnarray}
    \textit{MMD}(P_r,P_g) = \mathbb{E}_{x,x' \sim P_r} [k(x,x')] \,+\, \mathbb{E}_{y,y' \sim P_g} [k(y,y')] \,-\, 2 \mathbb{E}_{x\sim P_r,y \sim P_g} [k(x,y)],
    \label{eqn:MMD_Score}
\end{eqnarray}

where $P_r$ is the probability distribution of real images, $P_g$ is the probability distribution of generated images, and  $k(x,y)$ is a kernel function. From Eq.~\ref{eqn:MMD_Score}, when comparing generated images against real ones, one would expect low quality images to yield a high score and high quality images to yield a low score. As a kernel function we choose $k(x,y) = \vec{x}\cdot\vec{y}$, which only depends on the Euclidean distance between images in the two distributions. While there are potential issues with using a simple linear kernel on the raw images, we find that this metric strongly correlates with visual image quality and behaves as expected during training on the EBSD data set. To demonstrate this, we recorded the MMD score of the generated images at each epoch during the training of a Bernouilli WGAN model. The model was trained using 9000 ferrite images for 4000 epochs with  $\textit{$n_z$} = 16$ and $\lambda = 0.01$. 

\begin{figure}[H]
    \centering
    \includegraphics[width=0.85\linewidth]{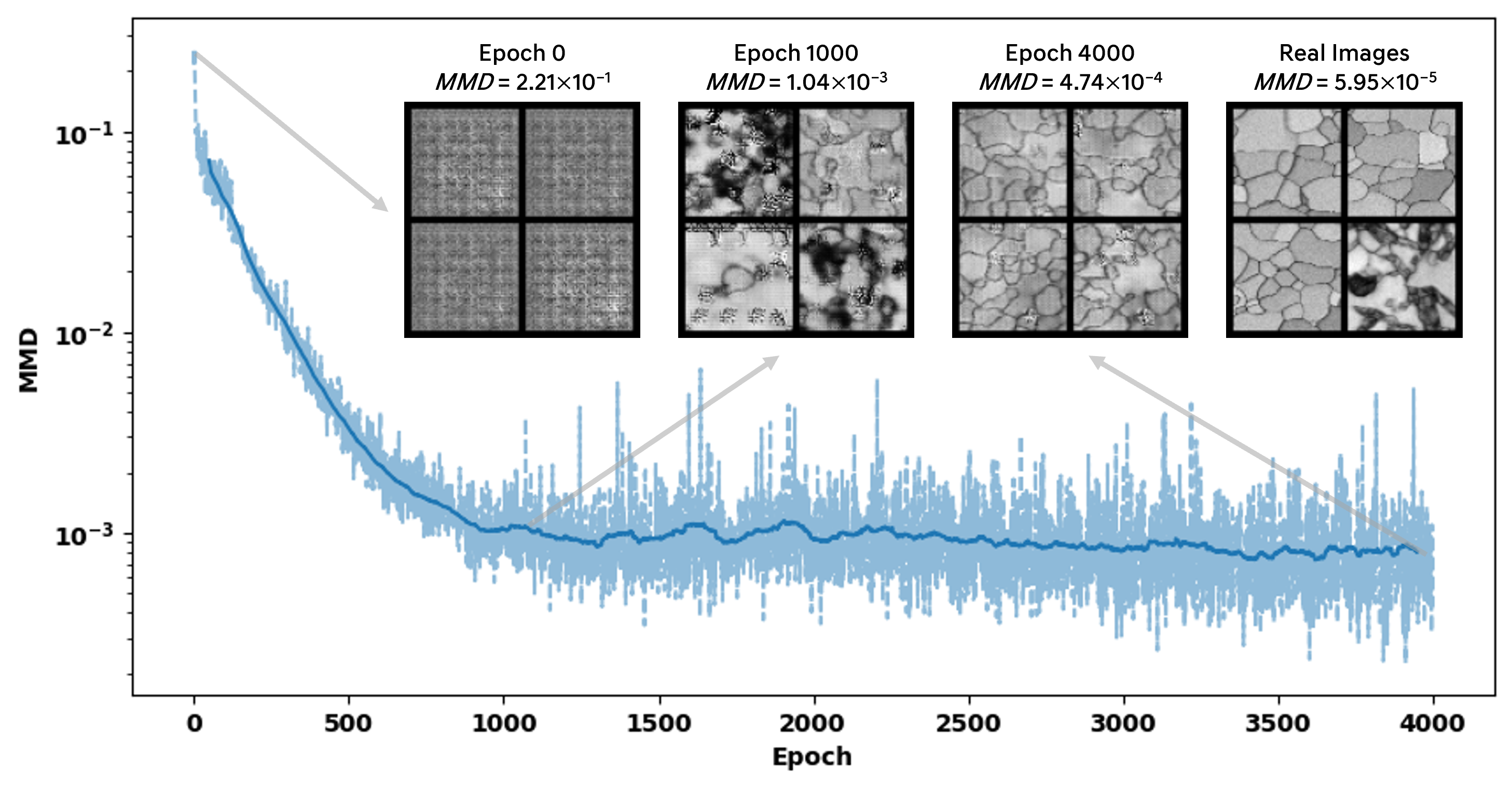}
    \caption{The MMD score of images generated by a Bernoulli WGAN model trained on ferrite images decreases during training and reflects the increased visual quality of the images generated. The MMD score of generated images at 0, 1000 and 4000 epochs is $2.21 \times 10^{-1}$, $1.04 \times 10^{-3}$, and $4.74 \times 10^{-4}$,  while the MMD score for real images is $6.0 \times 10^{-5} \pm 2 \times 10^{-5}$. The model 
 was trained with $\textit{$n_z$} = 16$ and $\lambda = 0.01$.}
    \label{fig:Bernoulli_MMD}
\end{figure}

In Fig.~\ref{fig:Bernoulli_MMD}, we see that the MMD score consistently decreases throughout the training, as expected from a model that learns the features of the images as training progresses. The quality of generated images, shown in insets for 0, 2000 and 4000 epochs, also improves throughout the training, as confirmed by a visual comparison with real images. That is, a lower MMD score appears to strongly correlate with visually higher image quality. We also find that the residual MMD score when comparing two finite size batches of real images to each other is $\approx 6\times10^{-5}$, which is lower than the best MMD score obtained from the generated images.  This demonstrates that the MMD metric can correctly measure image quality to the precision required in this work. Together, this evidence corroborates our assertion that the MMD score is a good measure of the quality of generated images. While this metric appears to be suitable for this work, in the future additional metrics may be needed to probe different characteristics of the generated images or to evaluate the image quality to a higher precision.

\subsection*{EBSD Data Preprocessing and Training Data}
EBSD samples were extracted from steel plates using wire electrical discharge machining.
The surfaces were ground with emery paper, followed by electro-polishing at $-20^{\circ}C$ using a solution of ethanol and perchloric acid in a 9:1 volume ratio.
EBSD mapping was performed using a JEOL JSM-7900F field emission scanning electron microscope equipped with an Oxford Instruments Symmetry EBSD detector.
The orientation mapping was conducted with step sizes ranging from 0.05 to 0.5 $\mu m$, at an acceleration voltage of 15 $kV$ and a probe current of 7.5 $nA$.
Individual electron backscatter diffraction patterns were collected at a resolution of 168 $\times$ 128 pixels with 8 $\times$ 8 binning, optimizing the balance between spatial resolution and data acquisition efficiency.

Raw EBSD datasets typically contain unindexed points or noise, often arising from grain boundaries, martensitic phases, surface defects, or equipment calibration errors.
To enhance data quality, a preprocessing algorithm based on a noisy pixel classification neural network was employed to mitigate these artifacts.
During the preprocessing step, noise caused by interfaces such as grain boundaries or scratches were corrected by interpolating orientation information from adjacent measurement points that exhibited the closest image quality (IQ) features.

A total of 20,000 individual grains were analyzed and each image patch was labeled according to its phase (\textit{i.e.}, ferrite or bainite) within the centrally located grain, providing 10,000 images for each phase.
The preprocessed EBSD data were segmented into 60 $\times$ 60 pixels image patches. The training dataset consists of five distinct channels.
The first channel, IQ, is a metric that primarily captures morphological features for microstructure phase classification.
The second channel encodes crystal structure using integer labels: 0 for noise, 1 for body-centered cubic, 2 for face-centered cubic, and 3 for other types of unit cells.
The third channel, kernel average misorientation (KAM), quantifies local misorientation and highlights geometrically dislocation densities.
In this channel, KAM values are assigned within the range of 0$^{\circ}$ to 3$^{\circ}$, and points affected by noise are assigned a default value of 3.5$^{\circ}$.
The fourth channel represents the deviation from the Kurdjumov-Sachs orientation relationship, providing information about displacive phase transformation, within the range of 0$^{\circ}$ to 15$^{\circ}$. Points where this measurement is not available are assigned a value of 20$^{\circ}$.
The fifth channel serves as a mask delineating the region of interest for main grain regions, with target areas marked as 1 and the background set to 0. All channels are linearly scaled within the range from 0 to 255.

\section*{Results} 
We now demonstrate that our hybrid quantum-classical GAN model can be trained to generate high quality synthetic samples of EBSD images of both the ferrite and bainite microstructures with improved image quality compared to the classical WGAN model alone. Implementing and optimizing the performance of a complex QML model requires a number of steps and iterations that cannot, at the present time, be exclusively executed on the quantum hardware alone. The typical development cycle involves quantitatively assessing the performance of the model by performing classical simulations of the quantum hardware. Through these simulations we provide statistically significant evidence that the QCBM-WGAN model provides improved performance, without including the additional complications present when running on the actual quantum hardware, such as gate noise and limitations on the number of circuit executions.

Only when the experiments produce a relevant result, the implementation is ported to the quantum hardware, where the final architecture and the best parameters from the noiseless experiments are adapted and tested. Often, intermediate ,noisy simulations, are performed from a selected number of noiseless experiments to refine the implementation and further validate the results before running the final test on the hardware. This step, nevertheless, requires the development of an accurate noise model that reproduces the specific characteristics of the hardware.

\subsection*{Ideal Simulations}
We implemented our quantum-classical GAN model by training the quantum circuit on the ideal, noiseless quantum simulator, and compared the performance of the resulting models with classical ``Bernouilli'' GANs trained in similar conditions. In all of our experiments, the WGAN parameter $\lambda$ was set to $0.01$, and the discriminator was trained at the same frequency as the generator.

\begin{figure}[H]
    \centering\includegraphics[width=0.9\linewidth]{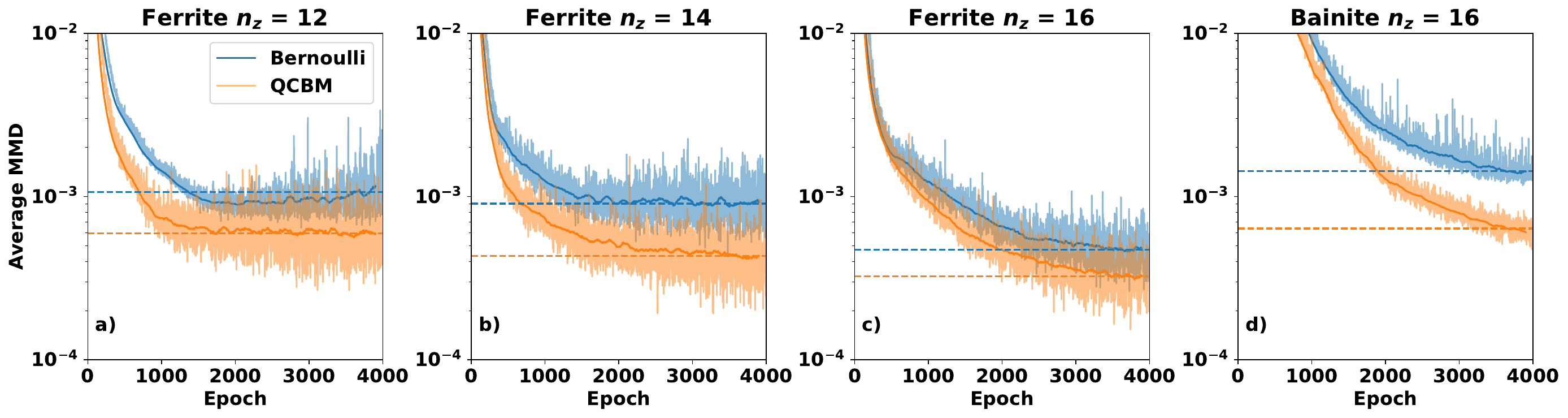}
    \caption{
    The average MMD score of the quantum models for the $n_z=12$, $14$ and $16$ outperform classical models of the same size. Ferrite models were trained on
    9000 images (a-c). Bainite models were trained on 3000 images (d). The averages are obtained from 5 independent runs in each case.}
\label{fig:mmd_scores_sim} 
\end{figure}

We trained distinct models in parallel for the two classes of materials, ferrite and bainite, given the substantial differences between the images in the two datasets.
The results show that ferrite (Fig.~\ref{fig:mmd_scores_sim}a-c) and bainite (Fig.~\ref{fig:mmd_scores_sim}d) models based on the QCBM-WGAN and trained on an ideal simulator outperform the classical Bernoulli models, obtaining lower final MMD score in all cases. 
The selected runs include ferrite models trained with $\textit{$n_z$} = 12$, $14$, and $16$ using 9000 images and a bainite model trained with $\textit{$n_z$} = 16$ using 3000 images.
In all cases the QCBM circuit was trained with $\alpha = 0.016$ and $\delta = 0.01$ for the first 100 epochs of the WGAN training and the  circuit parameters were fixed thereafter. The parameters were updated every 30 epochs for the training of the ferrite models and every 10 epochs for the bainite model. 
The data represent averages obtained over 5 independent runs in each case. Solid lines represent rolling averages over 100 epochs, and dashed horizontal lines highlight the final MMD score, averaged over the last 500 epochs. 

\begin{figure}[H]
    \centering    
    \includegraphics[width=0.92\linewidth]{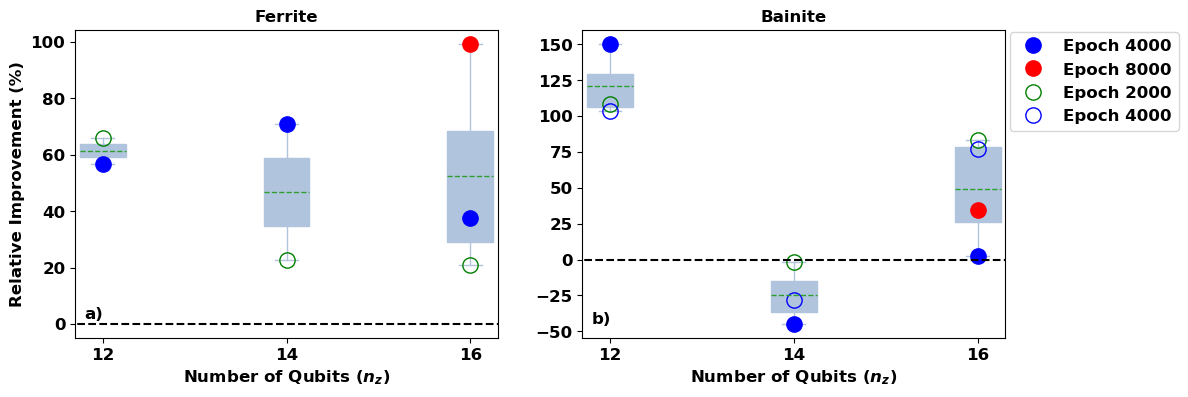}
    \caption{The overall relative improvement of (a) all ferrite models and (b) all bainite models tested for $\textit{$n_z$} = 12, 14$, and $16$. Solid markers are for models trained using 9000 images, and hollow markers are for models trained using 3000 images. The models were trained for 2000, 4000, and 8000 epochs. The dashed green line shows the mean relative improvement.}
    \label{fig:rel_improv} 
\end{figure}

The full exploration included further experiments, with more QCBM-based models trained to generate ferrite and bainite images using different values of $\textit{$n_z$}, \alpha$, number of images used for training, and number of training epochs. The performance of each resulting QCBM model was assessed by the degree of improvement relative to the correspondent Bernoulli model using:
\begin{equation}\label{eqn:5}
\textit{Relative Improvement} (\%)  = \frac{\textit{MMD}_B-\textit{MMD}_Q}{\frac{1}{2}(\textit{MMD}_B+\textit{MMD}_Q)}, 
\end{equation}

where $\textit{MMD}_B$ and $\textit{MMD}_Q$ are the MMD scores for the Bernoulli and QCBM models respectively, averaged over the final 100 training epochs. While the relative improvement between the quantum and classical showed significant statistical fluctuations between individual experiments, when taken together we see strong evidence for an overall improvement of the QCBM-WGAN model compared to the classical model. Fig.~\ref{fig:rel_improv} shows that the QCBM models outperformed Bernoulli models in all the ferrite cases, with relative improvement between 22\% and 99\%.
For the bainite models, the quantum model outperforms the classical model in 7 out of 10 cases and at least once for each value of $\textit{$n_z$}$, varying over a wider range between -45\% and 150\%. 
We recorded the best performances at $\textit{$n_z$} = 16$ and at $\textit{$n_z$} = 12$, respectively. This improvement in performance appears to persist to the largest quantum circuit size which we were able to simulate ($n_z=16$), indicating that continued improvement can expected in the future as the quantum circuit scales to sizes which are cannot be simulated classically. The results suggest that the quantum-enhanced implementation could offer striking advantage over the classical counterpart.

\subsection*{Simulations On A Quantum Processing Unit (QPU)}
The implementation processed in the noiseless simulator cannot be trivially ported to the current quantum devices and requires 
modifications and precautions.
In particular, it is critical to limit the number of function calls to the QPU to reduce the aggregate error due to quantum noise, and the number of 2-qubit gates in the QCBM circuit to improve fidelity.

\subsubsection*{Reduction of the number of hardware calls}
Focusing first on the number of function calls, the implementation we used in the noiseless case makes $2n$ calls to the quantum circuit per iteration to evaluate the gradient of the generator loss function (see Eq. \ref{eqn:Gloss_Grad}), where $n$ is the number of parameters in the QCBM circuit updated using gradient descent.
Given that the number of parameters scales quadratically with the number of qubits, one way to drastically reduce the total number of calls is to use a gradient-free optimizer to eliminate the dependency on the number of parameters. 
One method to optimize a system with multiple unknown parameters is the SPSA\cite{SPSA}.

Instead of evaluating gradients with respect to all parameters as in Eq. \ref{eqn:Gloss_Grad}, the SPSA optimizer adjusts the circuit parameters by sampling gradients along a single randomly selected direction. This cuts the number of calls to only two per iteration. Further reduction can be achieved by focusing on the initialization step, which requires $50$ additional calls to set the learning rate $a$, and perturbation $c$. 
By performing the initialization of the same circuit on the simulator with Aria-2 noise model it was possible to save the parameters to be used on the hardware runs.
The final number of hardware calls was therefore ultimately reduced to $2N + 1$ where $N$ is the number of SPSA iterations.


\subsubsection*{Reduction of the number of 2-qubit gates}
Quantum gates are unitary operators described by unitary matrices, and the matrix format for the $R_{xx}$ gates in the
QCBM circuit shown in Fig. \ref{fig:QCBM_WGAN}
is given by
\begin{equation}
R_{xx}(\theta) = CNOT(I\otimes R_x(\theta))CNOT,
\label{Rxx}
\end{equation}
where $CNOT$ is the matrix for the 2-qubit controlled NOT gate, $I$ is the identity matrix, and $R_x(\theta)$ is the matrix for the single qubit rotation gate \cite{MikeIke}. The $R_{xx}$ is not usually physically executed in most quantum hardware, and is rather implemented by using 2 $CNOT$ gates and single qubit rotations according to Eq.~\ref{Rxx}\cite{QiskitDecompose}. We apply the same decomposition in the QCBM circuit before its execution, but this operation results in an increase of the quantum error because the total number of 2-qubit gates is doubled.
A more efficient approach, aimed at limiting the 2-qubit error rate, is to convert the standard $R_x$, $R_z$, and $R_{xx}$ gates of the QCBM circuit into IonQ native gates, a set of quantum gates that are physically executed on IonQ's ion-trap computers. In this implementation, the $R_{xx}$ gates are directly converted into an equivalent circuit with 1- and 2-qubit native gates, keeping the overall number of 2-qubits the same.
The IonQ native gates are described in details in  \cite{NativeGates} and consist of the single qubit gates $\textit{GPI}(\phi)$, $\textit{GPI2}(\phi)$, $\textit{VirtualZ}(\theta)$, and the entangling gate $\textit{MS}(\phi_0,\phi_1)$. The $\textit{GPI}(\phi)$ gate can be considered a $\pi$ or bit-flip rotation with an embedded phase:

\begin{equation}
\textit{GPI}(\phi) = \begin{bmatrix}
            0 & e^{-i\phi}\\
            e^{i\phi} & 0
            \end{bmatrix}.
\end{equation}
The $\textit{GPI2}(\phi)$ gate could be considered an $\textit{RX}(\pi/2)$ or $\textit{RY}(\pi/2)$ with an embedded phase:
\begin{equation}
\textit{GPI2}(\phi) = \frac{1}{\sqrt{2}}\begin{bmatrix}
            1 & -ie^{-i\phi}\\
            -ie^{i\phi} & 1
            \end{bmatrix}.
\end{equation}

The $\textit{VirtualZ}(\theta)$ gate is equivalent to the standard $RZ(\theta)$:
\begin{equation}
\textit{VirtualZ}(\theta) = \frac{1} {\sqrt{2}}\begin{bmatrix}
            e^{-i\theta/2} & 0\\
            0 & e^{i\theta/2}
            \end{bmatrix}.
\end{equation}

The fully entangling $\textit{MS}(\phi_0,\phi_1)$ gate is an XX gate - a simultaneous, entangling $\pi/2$ rotation on both qubits:

\begin{equation}
\textit{MS}(\phi_0,\phi_1) = \frac{1} {\sqrt{2}}\begin{bmatrix}
            1 & 0 & 0 & -ie^{-i(\phi_0+\phi_1)}\\
            0 & 1 & -ie^{-i(\phi_0-\phi_1)} & 0\\
            0 & -ie^{i(\phi_0-\phi_1)} & 1 & 0 \\
            -ie^{i(\phi_0+\phi_1)} & 0 & 0 & 1
            \end{bmatrix}.
\end{equation}

The number of 2-qubit gates in the QCBM circuit can be further reduced by reducing the connectivity of the $R_{xx}$ gates. 
From the all-to-all connectivity of the $R_{xx}$ implemented in the noiseless experiments, referred to as the `full' version, we also developed a variation with connectivity limited to next three nearest neighbors that we define as the `reduced' version.


\subsubsection*{Selected experiments for the QPU}
For the final QPU training, we selected to run the models with 12 qubits QCBM ($\textit{$n_z$} = 12$) for both image classes, as all noiseless simulations in that case showed consistent improvements over the classical models.
In both cases the QCBM circuit was limited to 1 layer (see Fig. \ref{fig:QCBM_WGAN}), the number of training images to 3000, and the training epochs to 2000.


As an additional test, we chose to use the `full' version of the QCBM for the ferrite model, and the `reduced' version for the bainite model, resulting in 66 and 30 2-qubit gates respectively.
The SPSA optimization of the QCBM parameters was run for 50 iteration, requiring 101 circuit execution, every 50 epochs of the WGAN training.
Each optimization cycle was run 5 times for the ferrite model, to 250 training epochs, and 6 times for the bainite model, to 300 training epochs.

\begin{figure}[H]
    \centering
    \includegraphics[width=0.9\linewidth]{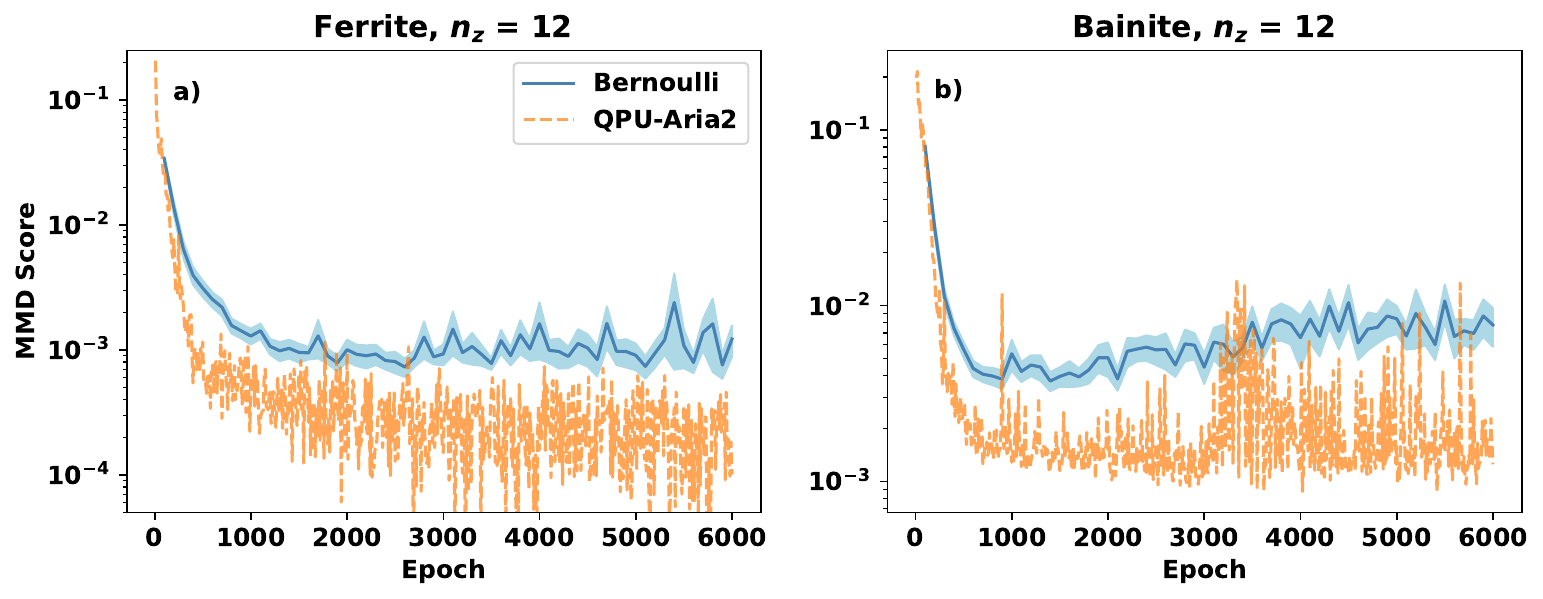}
    \caption{MMD scores during training for the classical models and the QCBM model trained using QPU backend. The ferrite models are compared in (a) and bainite images in (b). The MMD scores for the classical models are averages over 20 independent runs. The solid lines for the QCBM models are rolling averages over 150 epochs.}
    \label{fig:Ferrite_Bainite_mmd_qpu}
\end{figure}

Fig. \ref{fig:Ferrite_Bainite_mmd_qpu} shows that the quantum models train more effectively and reach a lower final MMD score than the classical models (averaged) for both ferrite and bainite images. The MMD scores for the Bernoulli models are averaged over 20 independent runs. The solid lines for the quantum models are rolling averages of 150 epochs.

\begin{figure}[H]
    \centering
    \includegraphics[width=0.9\linewidth]{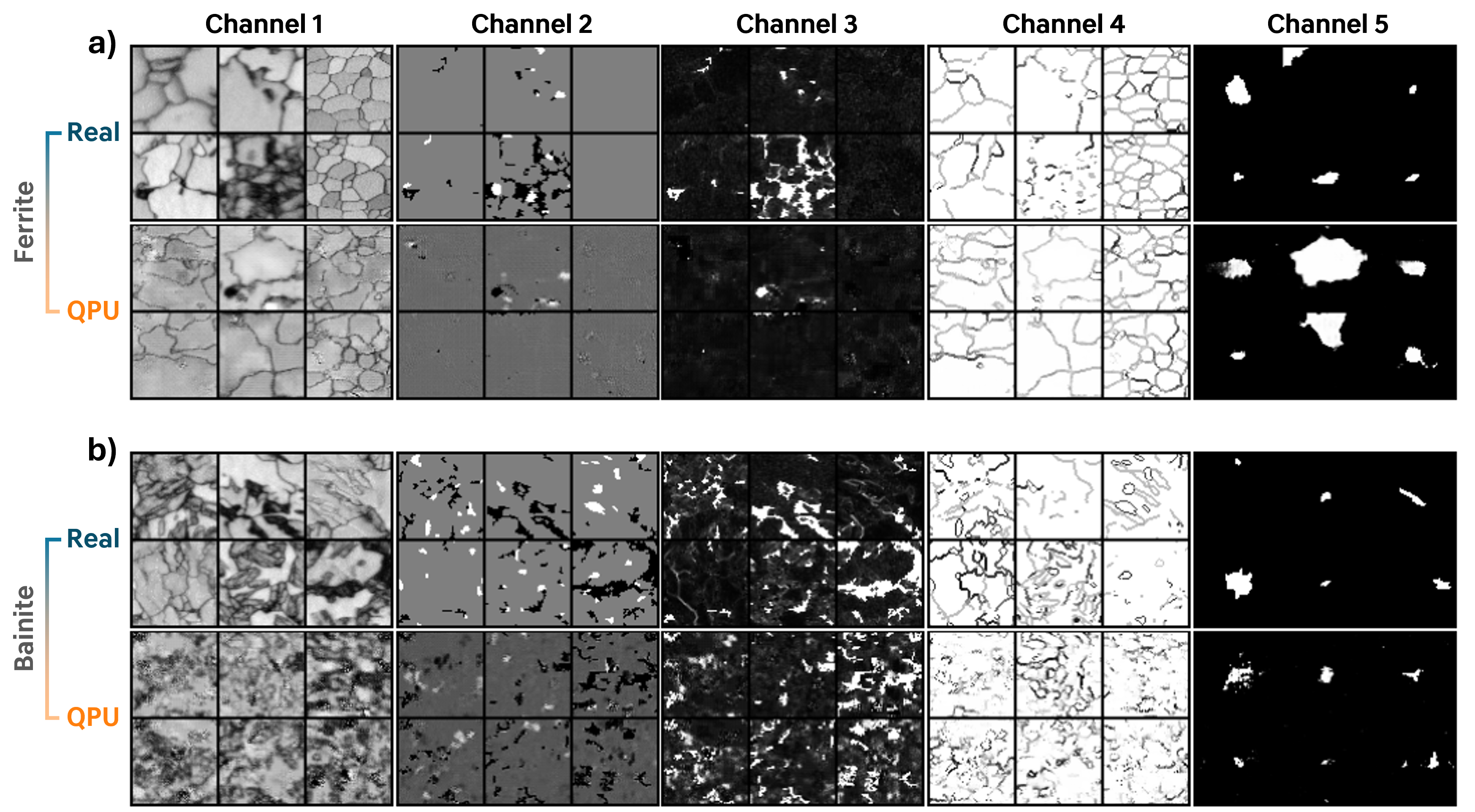}
    \caption{A channel by channel comparison between real images and images generated by the QCBM-WGAN model trained with the QPU backend. Ferrite comparison is shown in (a), and comparison for bainite images is shown in (b). Channels 1 to 5 represent IQ, types of unit cells, KAM, deviation from the K-S orientation relationship, and the mask of the region of interest, respectively. The final MMD scores for the QCBM ferrite and bainite models are $1.00\times10^{-4}$, and $1.25\times10^{-3}$ respectively.} 
    \label{fig:QPU_Ferr_Bain}
\end{figure}

Fig.~\ref{fig:QPU_Ferr_Bain}a shows that the QCBM model trained on the hardware is able to generate high quality ferrite images: the results are sharp and show clear contours, they reproduce the complex structures found in each of the channels of the real images, and present the varied output observed in the original dataset.

The QCBM model trained on the bainite dataset demonstrates a similar behavior as shown in Fig.~\ref{fig:QPU_Ferr_Bain}b. The generated images are tuned to the features of the different material class, reproducing characteristics that are distinct from the ferrite ones. Each set of generated images shows sharp details, and the correlation among all channels confirms that the model is able to learn the properties of the images.


In both cases
the QPU trained models are able to generate high quality images and outperform the corresponding classical models.
It should be noted that our efforts had been focused on the optimization of the ferrite model, which we studied extensively by tuning the training, exploring more parameters, and investigating its behaviors in more detail. The findings were then ported to the bainite class with only minor adjustments. It is therefore not surprising that the ferrite model achieved final MMD scores that are closer to the score of the real images compared to the bainite model. We believe, though, that the latter, with its different challenges in terms of image complexity and features, could reach the same performance with a dedicated study.

\section*{Conclusions}
We demonstrated that a ML model based on a WGAN that samples the output distribution of a quantum circuit can be trained to generate high quality synthetic 5-channel EBSD images. These synthetic images may provide significant value in supplementing sparse or difficult to obtain datasets of different steel microstructure images. We have provided strong evidence that by training a quantum circuit component at the input of this model we can consistently improve the performance over a state-of-the-art classical WGAN with the same structure. We further showed that it is feasible to perform the full end-to-end training of the hybrid model using current trapped ion quantum computers while maintaining this advantage. This is one of the first examples of a quantum generative model which is both applied to a highly complex real world data set and improves upon the performance of a modern classical WGAN model.

Using noiseless simulations of the quantum hardware to sample from the output of the quantum circuit, we trained hybrid quantum-classical models over a wide range of hyperparameters, where the dimension of latent vector, the number of images used for training, and the number of epochs for WGAN training were varied. We saw that overall, there is strong evidence that the hybrid quantum-classical model trains faster and generates images that reproduce more faithfully the features of the materials, with better definition and clarity than the classical model for two different classes of steel microstructures, ferrite and bainite. Both models were trained from limited image datasets and with a relatively small number of qubits encoding the latent vector.
Experiments showed relative improvements from the classical solution between 22\% and 99\% for the models used to generate ferrite images, and between -45\% and 150\% for the models used to generate bainite images, with positive relative improvements on 70\% of the cases.

The striking improvements of the hybrid quantum-classical models results on the noiseless simulator encouraged an end-to-end training of the quantum enhanced WGAN on the QPU. Despite the gate level noise on IonQ's Aria-2 QPU, the hybrid quantum-classical model reproduced the trends showed in the noiseless simulations, training faster and generating images with higher quality than its classical counterpart. These results show that our proposed quantum-enhanced WGAN model can be used to generate high quality synthetic complex images, improving on previous attempts to implement hybrid quantum-classical generative models on quantum annealers \cite{Benedetti_2018,vinci2019pathquantumadvantagetraining,Wilson_2021}, and NISQ devices \cite{PhysRevX.12.031010} that were focused on generating simpler images.
We believe that our work is the first demonstration of end-to-end training of quantum-classical generative models on a NISQ device that generate highly complex images with notable quality, outperforming their classical counterparts.

While our demonstration was based on only 12 qubits ($n_z = 12$), the implementation is scalable to even higher number of qubits. It will be a compelling future direction to explore training models with greater number of qubits that cannot be simulated classically. The high flexibility of our quantum-classical framework can be leveraged to generate a wider range
of diverse high quality outputs starting from even more complex datasets, extending results shown previously \cite{PhysRevApplied.21.044032}.

Potential future research directions include investigating the generalization capabilities of these hybrid quantum-classical models and exploring potential advantages of different quantum circuit architectures, including using matrix product state or other tensor network methods to classically pre-train the quantum circuit component \cite{2023APS..MARD73012M}.

An alterative area of exploration could be the integration of the QCBM-WGAN model into downstream applications. It would particularly interesting to study the effectiveness of using the synthetically generated images as data augmentation to train a classification model that can predict the microstructural properties from the crystallographic data. Further generalization would entail adapting the QCBM-WGAN model to generate more than two classes of images, which may be useful for more complex applications. Another direction to explore concerns methods for adding a quantum component to the discriminator of the GAN model, or using a stand-alone quantum classification model along with the augmented image set to classify the microstructure images.

Overall, we have effectively leveraged a QCBM based generative model to enhance the quality of synthetic EBSD images on both quantum simulators and quantum hardware. The promising results in this work highlight the potential use of quantum computers to enhance classical generative ML algorithms. Our results also indicate that there are many potential avenues to explore that may be able to realize near-term quantum advantages for problems related to materials science. The capability to generate physically accurate synthetic crystallographic images using a quantum-classical generative model may enhance image augmentation to mitigate data scarcity and streamline the preparation of datasets for robust high-throughput analysis. Further developing this approach while concurrently exploring more directions could boost quantum algorithms toward achieving practical quantum advantages in ML and related fields of research.

\end{document}